\documentclass{svmult}
\usepackage{epsfig,graphicx} % figures
\usepackage[bottom]{footmisc}% places footnotes at page bottom
\usepackage{natbib,url}      %RR additions
\bibpunct{(}{)}{;}{a}{}{,}   %RR reference punctuation
\def\cite#1{\citealp{#1}}    %RR restore old astroncite \cite command
\def\authorindex#1{}

\def\figspath{.}  %RA to be redefined by editor at insertion into book 

\def\ssr{Space\ Sci.\ Rev.,}
\begin{document}\newcount\preprintheader\preprintheader=1
%RR file: rr-latexdefs.tex = extra definitions by Rob Rutten
%RR last: Nov  5 2008 
%RR note: ?? problem    %RR Rob-to-Rob   

%RR ## to be adapted when the volume number is known
\def\thisvolume{these proceedings}

%RR journal abbreviations
%%%%%%%%%%%%%%%%%%%%%%%%%
\def\aj{{AJ}}			
\def\araa{{ARA\&A}}		
\def\apj{{ApJ}}			
\def\apjl{{ApJ}}		
\def\apjs{{ApJS}}		
\def\ao{{Appl.\ Optics}} 
\def\apss{{Ap\&SS}}		
\def\aap{{A\&A}}		
\def\aapr{{A\&A~Rev.}}		
\def\aaps{{A\&AS}}		
\def\an{{Astron.\ Nachrichten}}
\def\aspcs{{ASP Conf.\ Ser.}}
\def\azh{{AZh}}			
\def\baas{{BAAS}}		
\def\jrasc{{JRASC}}		
\def\memras{{MmRAS}}		
\def\mnras{{MNRAS}}
\def\nat{{Nat}}		
\def\pra{{Phys.\ Rev.\ A}} 
\def\prb{{Phys.\ Rev.\ B}}		
\def\prc{{Phys.\ Rev.\ C}}		
\def\prd{{Phys.\ Rev.\ D}}		
\def\prl{{Phys.\ Rev.\ Lett}}	
\def\pasp{{PASP}}
\def\pasj{{PASJ}}		
\def\qjras{{QJRAS}}
\def\science{{Sci}}		
\def\skytel{{S\&T}}		
\def\solphys{{Solar\ Phys.}} 
\def\sovast{{Soviet\ Ast.}}  
\def\ssr{{Space\ Sci.\ Rev.}}
\def\svassp{{Astrophys.\ Space Science Proc.}}
\def\zap{{ZAp}}			
\let\astap=\aap
\let\apjlett=\apjl
\let\apjsupp=\apjs

%RR astronomy and math commands copied from ASP
%%%%%%%%%%%%%%%%%%%%%%%%%%%%%%%%%%%%%%%%%%%%%%%
\def\ion#1#2{{\rm #1}\,{\uppercase{#2}}}  %RR ~>\, \sc > uc 
\def\deg{\hbox{$^\circ$}}
\def\sun{\hbox{$\odot$}}
\def\earth{\hbox{$\oplus$}}
\def\la{\mathrel{\hbox{\rlap{\hbox{\lower4pt\hbox{$\sim$}}}\hbox{$<$}}}}
\def\ga{\mathrel{\hbox{\rlap{\hbox{\lower4pt\hbox{$\sim$}}}\hbox{$>$}}}}
\def\sq{\hbox{\rlap{$\sqcap$}$\sqcup$}}
\def\arcmin{\hbox{$^\prime$}}
\def\arcsec{\hbox{$^{\prime\prime}$}}
\def\fd{\hbox{$.\!\!^{\rm d}$}}
\def\fh{\hbox{$.\!\!^{\rm h}$}}
\def\fm{\hbox{$.\!\!^{\rm m}$}}
\def\fs{\hbox{$.\!\!^{\rm s}$}}
\def\fdg{\hbox{$.\!\!^\circ$}}
\def\farcm{\hbox{$.\mkern-4mu^\prime$}}
\def\farcs{\hbox{$.\!\!^{\prime\prime}$}}
\def\fp{\hbox{$.\!\!^{\scriptscriptstyle\rm p}$}}
\def\micron{\hbox{$\mu$m}}
\def\onehalf{\hbox{$\,^1\!/_2$}}	
\def\onethird{\hbox{$\,^1\!/_3$}}
\def\twothirds{\hbox{$\,^2\!/_3$}}
\def\onequarter{\hbox{$\,^1\!/_4$}}
\def\threequarters{\hbox{$\,^3\!/_4$}}
\def\ubv{\hbox{$U\!BV$}}		
\def\ubvr{\hbox{$U\!BV\!R$}}		
\def\ubvri{\hbox{$U\!BV\!RI$}}		
\def\ubvrij{\hbox{$U\!BV\!RI\!J$}}		
\def\ubvrijh{\hbox{$U\!BV\!RI\!J\!H$}}		
\def\ubvrijhk{\hbox{$U\!BV\!RI\!J\!H\!K$}}		
\def\ub{\hbox{$U\!-\!B$}}		
\def\bv{\hbox{$B\!-\!V$}}		
\def\vr{\hbox{$V\!-\!R$}}		
\def\ur{\hbox{$U\!-\!R$}}

%%%%%%%%%%%%%%%%%%%%%%%%%%%%%%%%%%%%%%%%%%%%%%%%%%%%%%%%%%%%%%%%%%%%%%%%%%%%
%RR RJR additional commands
%%%%%%%%%%%%%%%%%%%%%%%%%%%%%%%%%%%%%%%%%%%%%%%%%%%%%%%%%%%%%%%%%%%%%%%%%%%%

%RR -- non-bullet item marker in itemize list 
\def\labelitemi{{\bf --}}  

%RR -- latin abbreviations
\def\rmit#1{{\it #1}}              %% italics (RR style, Kluwer)
\def\rmit#1{{\rm #1}}              %% redefine for ASP, A&A, ApJ, Springer??
\def\etal{\rmit{et al.}}           %% use \etal\ for space behind it        
\def\etc{\rmit{etc.}}           
\def\ie{\rmit{i.e.,}}              %% , required (Webster 1681)
\def\eg{\rmit{e.g.,}}              %% , required (Webster 1681)
\def\cf{cf.}                       %% no Latin, always Roman (Webster 1686)
\def\viz{\rmit{viz.}}
\def\vs{\rmit{vs.}}

%RR -- mathematical
\def\rot{\hbox{\rm rot}}
\def\div{\hbox{\rm div}}
\def\lesssim{\mathrel{\hbox{\rlap{\hbox{\lower4pt\hbox{$\sim$}}}\hbox{$<$}}}}
\def\gtrsim{\mathrel{\hbox{\rlap{\hbox{\lower4pt\hbox{$\sim$}}}\hbox{$>$}}}}
\def\dif{\: {\rm d}}                       %% differential d with space
\def\ep{\:{\rm e}^}                        %% e^ with space and roman e
\def\dash{\hbox{$\,-\,$}}                  %% math-like hyphen
\def\is{\!=\!}                             %% = in text for tighter spacing

%RR --stellar stuff
\def\starname#1#2{${#1}$\,{\rm {#2}}}  %% \starname{\alpha}{Cen~A} 
\def\Teff{\hbox{$T_{\rm eff}$}}   

%RR -- units (in addition to the ASP ones above)
\def\kms{\hbox{km$\;$s$^{-1}$}}
\def\Mxcm{\hbox{Mx\,cm$^{-2}$}}    %% no 2, damn tex

%RR -- magnetic field 
\def\Bapp{\hbox{$B_{\rm app}$}}    %% apparent flux density, Lites convention

%RR -- oscillations
\def\komega{($k, \omega$)}                 %% k - omega 
\def\kf{($k_h,f$)}                         %% f - k_h
\def\VminI{\hbox{$V\!\!-\!\!I$}}           %% V-I
\def\IminI{\hbox{$I\!\!-\!\!I$}}           %% I-I
\def\VminV{\hbox{$V\!\!-\!\!V$}}           %% V-V
\def\Xt{\hbox{$X\!\!-\!t$}}                %% X-t

%RR -- atomic levels
%%      use:    \level 3s3p 3Pe
%%              \level 3s$^2$ {1,3}P{e,o}
%%              \level {} 3Ge
\def\level #1 #2#3#4{$#1 \: ^{#2} \mbox{#3} ^{#4}$}   

%RR -- some spectral species
\def\specchar#1{\uppercase{#1}}    %% to be redefined for A&A = \sc
\def\AlI{\mbox{Al\,\specchar{i}}}  %% use \AlI\ for space behind it
\def\BI{\mbox{B\,\specchar{i}}} 
\def\BII{\mbox{B\,\specchar{ii}}}  
\def\BaI{\mbox{Ba\,\specchar{i}}}  
\def\BaII{\mbox{Ba\,\specchar{ii}}} 
\def\CI{\mbox{C\,\specchar{i}}} 
\def\CII{\mbox{C\,\specchar{ii}}} 
\def\CIII{\mbox{C\,\specchar{iii}}} 
\def\CIV{\mbox{C\,\specchar{iv}}} 
\def\CaI{\mbox{Ca\,\specchar{i}}} 
\def\CaII{\mbox{Ca\,\specchar{ii}}} 
\def\CaIII{\mbox{Ca\,\specchar{iii}}} 
\def\CoI{\mbox{Co\,\specchar{i}}} 
\def\CrI{\mbox{Cr\,\specchar{i}}} 
\def\CriI{\mbox{Cr\,\specchar{ii}}} 
\def\CsI{\mbox{Cs\,\specchar{i}}} 
\def\CsII{\mbox{Cs\,\specchar{ii}}} 
\def\CuI{\mbox{Cu\,\specchar{i}}} 
\def\FeI{\mbox{Fe\,\specchar{i}}} 
\def\FeII{\mbox{Fe\,\specchar{ii}}} 
\def\FeIX{\mbox{Fe\,\specchar{ix}}}
\def\FeX{\mbox{Fe\,\specchar{x}}}
\def\FeXVI{\mbox{Fe\,\specchar{xvi}}}
\def\FrI{\mbox{Fr\,\specchar{i}}}
\def\HI{\mbox{H\,\specchar{i}}} 
\def\HII{\mbox{H\,\specchar{ii}}} 
\def\Hmin{\hbox{\rmH$^{^{_{\scriptstyle -}}}$}}      %% H^min, elegant
\def\Hemin{\hbox{{\rm He}$^{^{_{\scriptstyle -}}}$}} %% He^min, idem
\def\HeI{\mbox{He\,\specchar{i}}} 
\def\HeII{\mbox{He\,\specchar{ii}}} 
\def\HeIII{\mbox{He\,\specchar{iii}}} 
\def\KI{\mbox{K\,\specchar{i}}} 
\def\KII{\mbox{K\,\specchar{ii}}} 
\def\KIII{\mbox{K\,\specchar{iii}}} 
\def\LiI{\mbox{Li\,\specchar{i}}} 
\def\LiII{\mbox{Li\,\specchar{ii}}} 
\def\LiIII{\mbox{Li\,\specchar{iii}}} 
\def\MgI{\mbox{Mg\,\specchar{i}}} 
\def\MgII{\mbox{Mg\,\specchar{ii}}} 
\def\MgIII{\mbox{Mg\,\specchar{iii}}} 
\def\MnI{\mbox{Mn\,\specchar{i}}} 
\def\NI{\mbox{N\,\specchar{i}}}
\def\NaI{\mbox{Na\,\specchar{i}}}
\def\NaII{\mbox{Na\,\specchar{ii}}}
\def\NaIII{\mbox{Na\,\specchar{iii}}} 
\def\NiI{\mbox{Ni\,\specchar{i}}} 
\def\NiII{\mbox{Ni\,\specchar{ii}}}
\def\NiIII{\mbox{Ni\,\specchar{iii}}} 
\def\OI{\mbox{O\,\specchar{i}}} 
\def\OVI{\mbox{O\,\specchar{vi}}}
\def\RbI{\mbox{Rb\,\specchar{i}}} 
\def\SII{\mbox{S\,\specchar{ii}}} 
\def\SiI{\mbox{Si\,\specchar{i}}} 
\def\SiII{\mbox{Si\,\specchar{ii}}} 
\def\SrI{\mbox{Sr\,\specchar{i}}}
\def\SrII{\mbox{Sr\,\specchar{ii}}}
\def\TiI{\mbox{Ti\,\specchar{i}}} 
\def\TiII{\mbox{Ti\,\specchar{ii}}} 
\def\TiIII{\mbox{Ti\,\specchar{iii}}} 
\def\TiIV{\mbox{Ti\,\specchar{iv}}} 
\def\VI{\mbox{V\,\specchar{i}}} 
\def\HtwoO{\mbox{H$_2$O}}        %% H2O %RR TeX doesn't accept numbers alas
\def\Otwo{\mbox{O$_2$}}          %% O2

%RR -- hydrogen spectrum features
\def\Halpha{\mbox{H\hspace{0.1ex}$\alpha$}} %% \Halpha\ for space behind it
\def\Ha{\mbox{H\hspace{0.2ex}$\alpha$}}
\def\Hbeta{\mbox{H\hspace{0.2ex}$\beta$}}
\def\Hgamma{\mbox{H\hspace{0.2ex}$\gamma$}}
\def\Hdelta{\mbox{H\hspace{0.2ex}$\delta$}}
\def\Hepsilon{\mbox{H\hspace{0.2ex}$\epsilon$}}
\def\Hzeta{\mbox{H\hspace{0.2ex}$\zeta$}}
\def\Lyalpha{\mbox{Ly$\hspace{0.2ex}\alpha$}}
\def\Lybeta{\mbox{Ly$\hspace{0.2ex}\beta$}}
\def\Lygamma{\mbox{Ly$\hspace{0.2ex}\gamma$}}
\def\Lycont{\mbox{Ly\hspace{0.2ex}{\small cont}}}
\def\Baalpha{\mbox{Ba$\hspace{0.2ex}\alpha$}}
\def\Babeta{\mbox{Ba$\hspace{0.2ex}\beta$}}
\def\Bacont{\mbox{Ba\hspace{0.2ex}{\small cont}}}
\def\Paalpha{\mbox{Pa$\hspace{0.2ex}\alpha$}}
\def\Bralpha{\mbox{Br$\hspace{0.2ex}\alpha$}}

%RR -- Na D
\def\NaD{\mbox{Na\,\specchar{i}\,D}}    %% use \NaD\ for space behind it
\def\NaDone{\mbox{Na\,\specchar{i}\,\,D$_1$}}
\def\NaDtwo{\mbox{Na\,\specchar{i}\,\,D$_2$}}
\def\NaID{\mbox{Na\,\specchar{i}\,\,D}}
\def\NaIDone{\mbox{Na\,\specchar{i}\,\,D$_1$}}
\def\NaIDtwo{\mbox{Na\,\specchar{i}\,\,D$_2$}}
\def\Done{\mbox{D$_1$}}
\def\Dtwo{\mbox{D$_2$}}

%RR -- Mg b 
\def\Mgbone{\mbox{Mg\,\specchar{i}\,b$_1$}}
\def\Mgbtwo{\mbox{Mg\,\specchar{i}\,b$_2$}}
\def\Mgbthree{\mbox{Mg\,\specchar{i}\,b$_3$}}
\def\MgIb{\mbox{Mg\,\specchar{i}\,b}}
\def\MgIbone{\mbox{Mg\,\specchar{i}\,b$_1$}}
\def\MgIbtwo{\mbox{Mg\,\specchar{i}\,b$_2$}}
\def\MgIbthree{\mbox{Mg\,\specchar{i}\,b$_3$}}

%RR -- Ca II H & K 
\def\CaIIK{\mbox{Ca\,\specchar{ii}\,K}}       %% use \CaIIK\ for space
\def\CaIIH{\mbox{Ca\,\specchar{ii}\,H}}
\def\CaIIHK{\mbox{Ca\,\specchar{ii}\,H\,\&\,K}}
\def\HK{\mbox{H\,\&\,K}}
\def\Kthree{\mbox{K$_3$}}      %% numbers not permitted, alas
\def\Hthree{\mbox{H$_3$}}
\def\Ktwo{\mbox{K$_2$}}
\def\Htwo{\mbox{H$_2$}}
\def\Kone{\mbox{K$_1$}}     
\def\Hone{\mbox{H$_1$}}     
\def\KtwoV{\mbox{K$_{2V}$}}
\def\KtwoR{\mbox{K$_{2R}$}}
\def\KoneV{\mbox{K$_{1V}$}}
\def\KoneR{\mbox{K$_{1R}$}}
\def\HtwoV{\mbox{H$_{2V}$}}
\def\HtwoR{\mbox{H$_{2R}$}}
\def\HoneV{\mbox{H$_{1V}$}}
\def\HoneR{\mbox{H$_{1R}$}}

%RR -- Mg II h & k 
\def\hk{\mbox{h\,\&\,k}}
\def\kthree{\mbox{k$_3$}}    
\def\hthree{\mbox{h$_3$}}
\def\ktwo{\mbox{k$_2$}}
\def\htwo{\mbox{h$_2$}}
\def\kone{\mbox{k$_1$}}     
\def\hone{\mbox{h$_1$}}     
\def\ktwoV{\mbox{k$_{2V}$}}
\def\ktwoR{\mbox{k$_{2R}$}}
\def\koneV{\mbox{k$_{1V}$}}
\def\koneR{\mbox{k$_{1R}$}}
\def\htwoV{\mbox{h$_{2V}$}}
\def\htwoR{\mbox{h$_{2R}$}}
\def\honeV{\mbox{h$_{1V}$}}
\def\honeR{\mbox{h$_{1R}$}}

\title*{Interplanetary Consequences of a Large CME}
\titlerunning{Interplanetary Consequences of a Large CME}
%Intense Flare/CME event: Interplanetary Consequences }  

\author {M. Lahkar\inst{1,2}
        \and
        P. K. Manoharan\inst{2}
         \and      
        K. Mahalakshmi\inst{2}
        \and
        K. Prabhu\inst{2} 
	\and  \\
	G. Agalya\inst{2}
	\and
        S. Shaheda Begum\inst{2}
	\and
	P. Revathi\inst{2}
	}

\authorindex{Lahkar, M.}
\authorindex{Manoharan, P. K.}
\authorindex{Mahalakshmi, K.}
\authorindex{Prabhu, K.}
\authorindex{Agalya, G.}
\authorindex{Shaheda Begum, S.}
%RR?? above name indeed double?
\authorindex{Revathi, P.}

\authorrunning{Lahkar et al.}  
\institute{ Centre for Radio Astronomy, Department of Physics, \\
            Cotton College, Guwahati, Assam, India
\and
            Radio Astronomy Centre, National Centre for Radio Astrophysics,\\
            Tata Institute of Fundamental Research, Udhagamandalam (Ooty), India}
\maketitle
\setcounter{footnote}{0} 

\begin{abstract} 

We analyze a coronal mass ejection (CME) which resulted from an
intense flare in active region AR486 on November 4, 2003.  The CME
propagation and speed are studied with interplanetary scintillation
images, near-Earth space mission data, and Ulysses
measurements. Together, these diverse diagnostics suggest that the
internal magnetic energy of the CME determines its interplanetary
consequences.

\end{abstract}

\begin{figure}[b] 
\centering
\includegraphics[width=35mm]{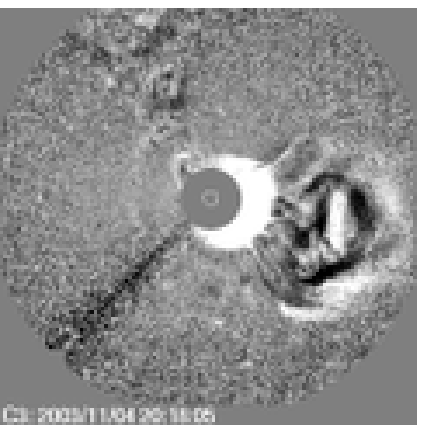}
\includegraphics[width=35mm]{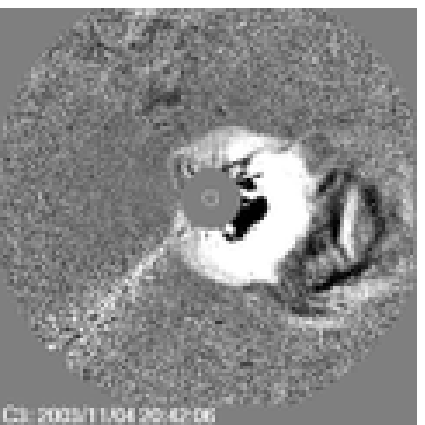}
\includegraphics[width=35mm]{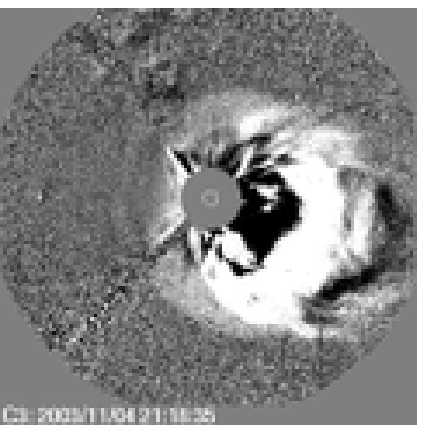}
\caption[]{\label{lahkar-fig:fig3a}
LASCO images showing CME--CME interaction
}
\end{figure}

\section{Introduction: intense flare and CME}

An intense flare and associated coronal mass ejection (CME) occurred
on November 4, 2003, during 19:50--20:10~UT in active region AR486.
The onset of the halo CME was observed in the SOHO/LASCO C2 field of view
(\cite{lahkar-1995SoPh..162..357B}) at 19:54~UT. The CME propagated
rather fast in the C2--C3 field of view, at about 2650~\kms\ linear
speed (Fig.~\ref{lahkar-fig:fig3a}), and caught up with a preceding
CME that originated at the same location at 12:54~UT but had much lower
linear speed (about 600~\kms) and narrow sky-plane width (about
70$^\circ$). The LASCO images show interaction of the two CMEs at about
25$R_\odot$.  It caused remarkably complex and intense radio emission
(\eg\ \cite{lahkar-2001ApJ...548L..91G}).

\section{Radio spectra, particle fluxes, and scintillation images}
 
Prior to the CME-CME interaction, intense type-III bursts were
observed with space missions Wind, Cassini
(\cite{lahkar-2004SSRv..115....1K}), 
and Ulysses. In the WAVES spectrum 
(\cite{lahkar-1995SSRv...71..231B}), 
a fast-drifting intense
type-II burst was observed during 19:50--21:00~UT at frequencies
1--11~MHz.  It arrived at 1~MHz around 20:45~UT, corresponding to 
$R = 10 -15 R_\odot$ in good agreement with the LASCO data.

\begin{figure} 
\centering
\includegraphics[width=10cm]{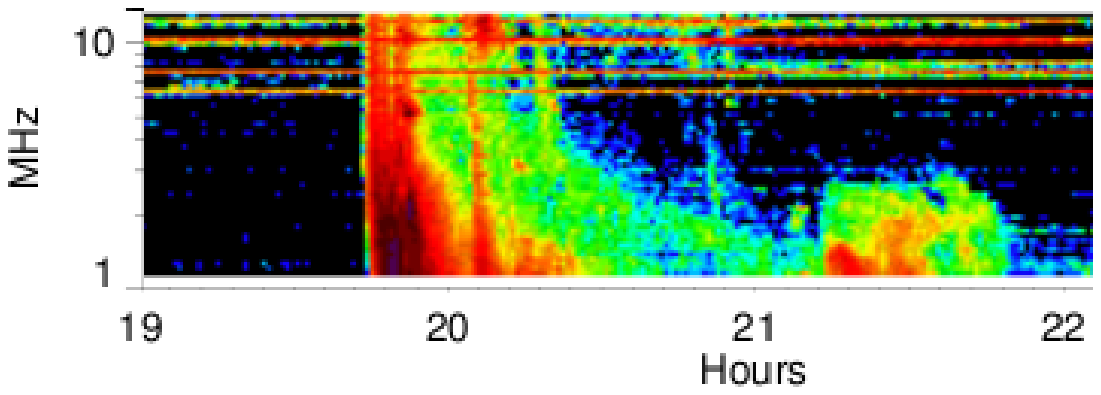}\\
\includegraphics[width=10cm]{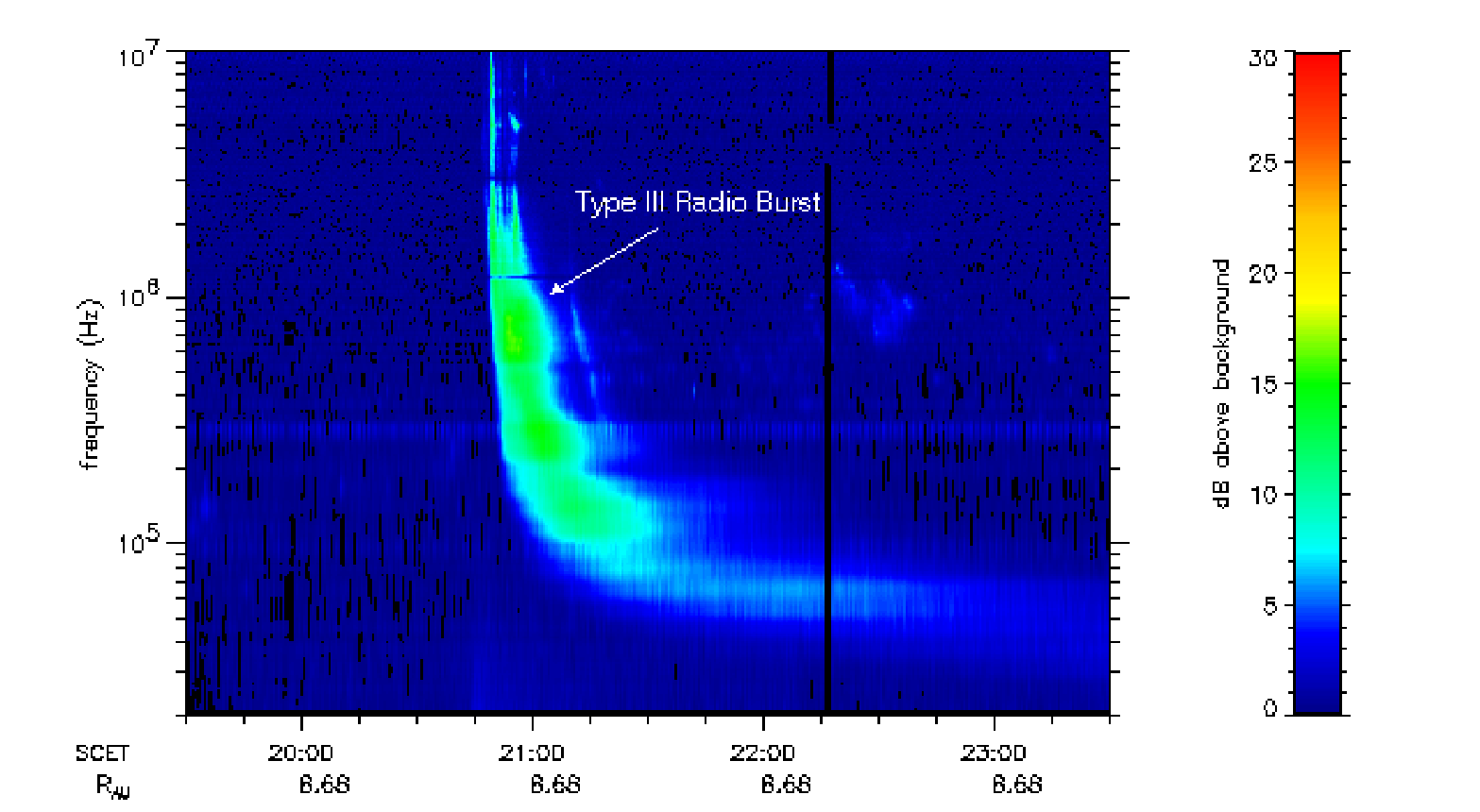}
\caption[]{\label{lahkar-fig:fig4a}
Radio spectra from the WAVES and Cassini missions. Cassini was 
located at 8.7~AU. 
}
\end{figure}

The WAVES spectrum shows emission from the CME--CME interaction in the
frequency range 1--3~MHz, higher than the type-II frequency at that
time. Thus, this emission relates to about 15 times higher density
than the typical ambient density at the interaction height (\eg\
\cite{lahkar-2001ApJ...548L..91G}).  In the Cassini spectrum, an
intense patch of emission occurred at frequencies below 1 MHz as the
extension of the complex emission seen in WAVES.  Thus, the
interaction led to electron acceleration through the intense magnetic
field and reconnection resulting from the interaction. The interaction
and associated phenomena were observed more than 30 minutes on the
above radio spectra (Fig.~\ref{lahkar-fig:fig4a}).

Type-IV emission was observed at frequencies $>7$~MHz during 
20:10--21:00 UT (WAVES spectrum in Fig.~\ref{lahkar-fig:fig4a})
suggesting a plasmoid associated with the 
fast-moving CME (\eg\ \cite{lahkar-2003ApJ...592..597M}).  
However, it disappeared just before the interaction.

\begin{figure}
\centering
\includegraphics[width=\textwidth]{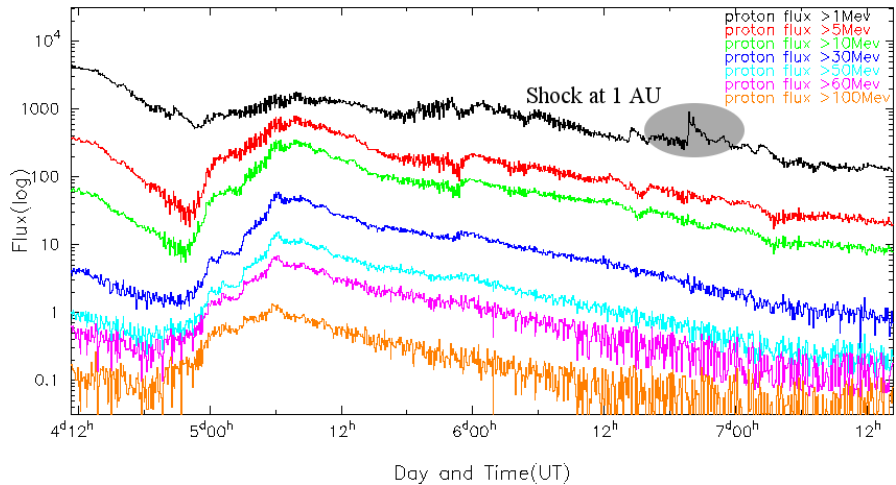}
\caption[]{\label{lahkar-fig:fig5}
  Particle profiles in different energy bands. 
}
\end{figure}

The particle flux in the energy range 1--100~MeV showed
no enhancement during the flare,
but around 21:30~UT it increased in all energy bands to peak at 
about 06:00~UT the next day.  Figure~\ref{lahkar-fig:fig5}
shows that the increase was about 50 times above the 
pre-flare value at energies above 5~MeV.  The sudden 
increase suggests that the CME-CME interaction favoured development
of magnetic connectivity with the Earth and particle acceleration
(\eg\ \cite{lahkar-2006AGUFMSH43B1518G}).
The CME-produced shock arrived on November 6, 19:20~UT
(1-MeV profile in Fig.~\ref{lahkar-fig:fig5}).

\begin{figure}[t]
\centering
\includegraphics[width=12cm]{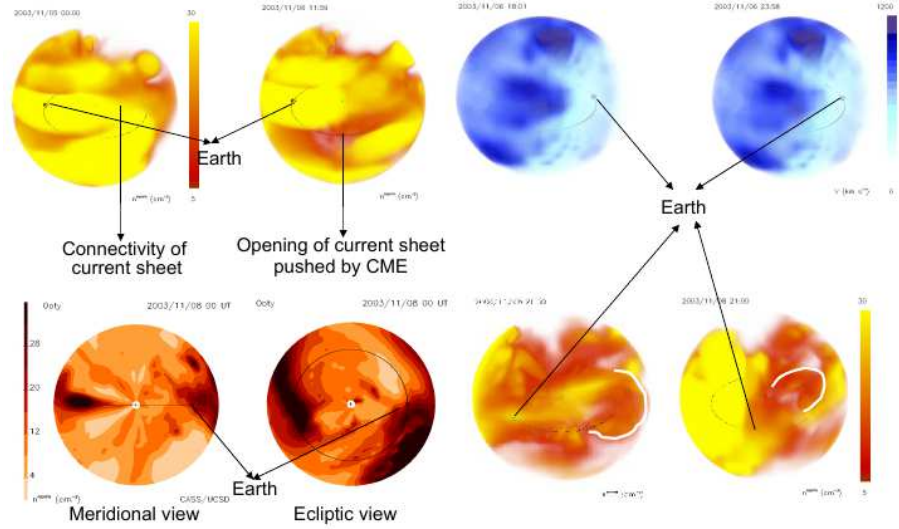}
\caption[]{\label{lahkar-fig:fig6}
  Ooty IPS images showing the current sheet location (top left), CME
  deflection by the coronal hole (top right), CME compression of the
  solar wind (bottom left), and CME propagation (bottom right). The
  Sun is located at the center of each image. The two images at top 
  right represent solar-wind speed; the others represent density.  }
\end{figure}

Figure~\ref{lahkar-fig:fig6} shows 3-d tomographic interplanetary
scintillation (IPS) images of the CME obtained with the Ooty Radio
Telescope (\eg\ \cite{lahkar-2001ApJ...559.1180M}).  They cover a
range of 50--250~$R_\odot$.  The interacting CMEs compressed the
high-density and low-speed solar wind originating above a current
sheet along the North-South direction.  The IPS images obtained during
November 6, 18:00--24:00~UT show the pushing and opening of the
current sheet as viewed from the Earth.  The central part of the CME
compressed the fast solar wind belonging to a large coronal hole,
which deflected the CME further away from the Sun-Earth line.

\begin{figure}
\centering
\includegraphics[angle=-90,width=58mm]{\figspath/lahkar-fig7a}
\includegraphics[angle=-90,width=58mm]{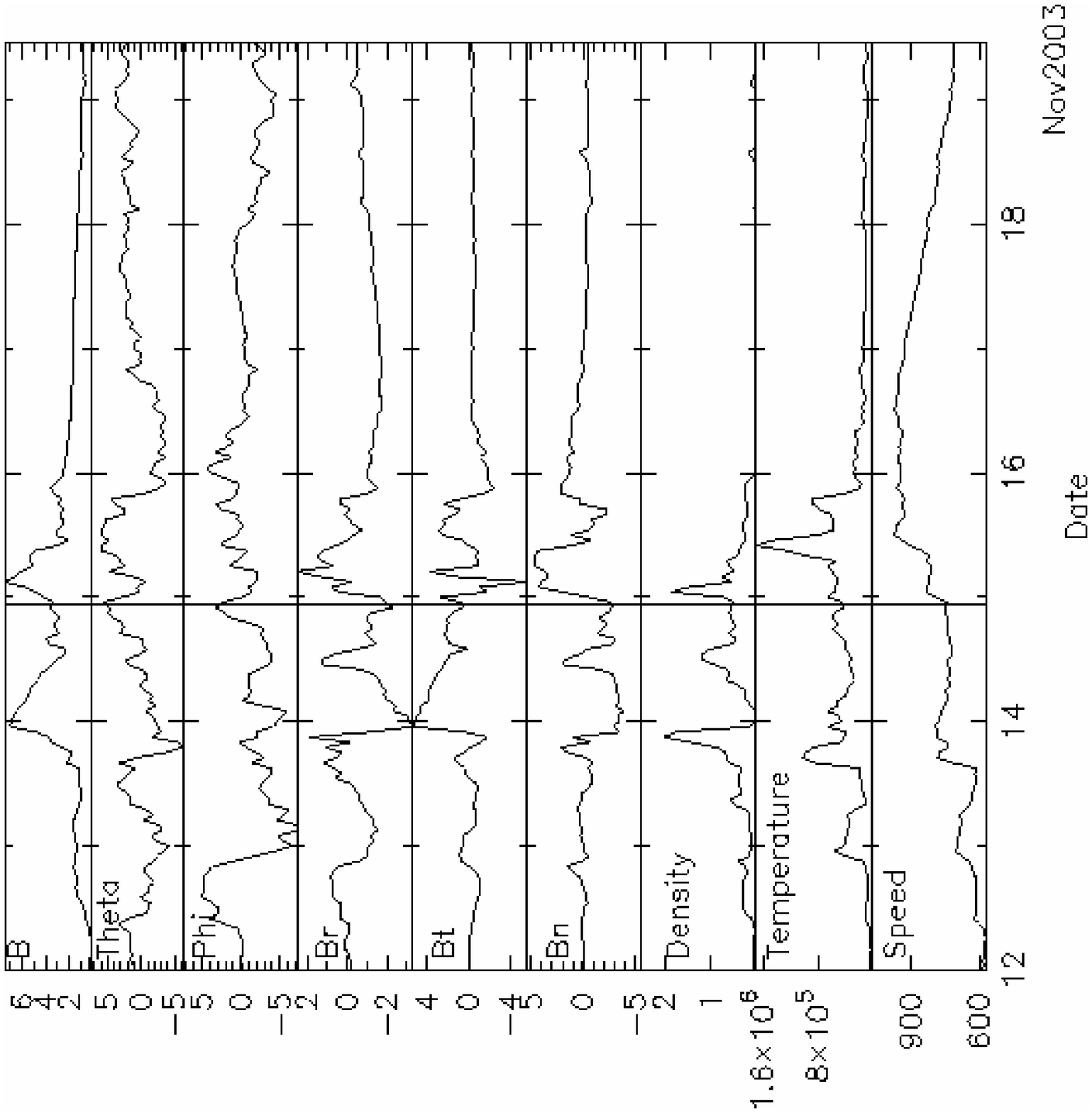}
\caption[]{\label{lahkar-fig:fig7a}
1-AU and  Ulysses hourly averages of solar wind 
         parameters.
}
\end{figure}

The solar wind parameters observed by near-Earth and with the Ulysses
space missions are shown in Fig.~\ref{lahkar-fig:fig7a}.  The shock
arrival times are shown by vertical lines. Since Ulysses was favourably
located in the CME propagation direction, it could record the nose
part of the CME and its shock, as indicated by a speed value of over
900~\kms\ at 5 AU.  At Earth, the shock speed was below 600~\kms,
suggesting that the eastern tail swept the Earth.  From these
measurements we infer a speed profile $V \sim R^{-0.4}$ to Earth.
However, the deceleration $V \sim R^{-0.2}$ out to 5~AU near Ulysses 
implies gradual decline in speed along the CME propagation direction, 
which is in good agreement with the IPS measurements.

\section{Conclusion}

Our study shows the characteristics of a fast-moving CME and its
interactions with transient and solar-wind structures at different
distances from the Sun with good consistency between diverse
diagnostics.  The enhancement in radio emission and production of
high-energy particles suggest that the magnetic field associated with
the CME was strong.  The gradual decline in CME speed suggests that
the internal magnetic energy of the CME supported its propagation
including expansion in overcoming the aerodynamical drag imposed by
the ambient solar wind (\eg\
\cite{lahkar-2006SoPh..235..345M}).  

\begin{acknowledgement}
We thank the Cassini, GOES, SOHO, TRACE, Ulysses, Wind, and
OMNI-database teams for making their data available on the web.  We
also thank B.~Jackson and the UCSD team for the IPS tomography analysis
package.  M. Lahkar thanks the National Centre for Radio Astrophysics
(TIFR) for financial support. This work is partially
supported by the CAWSES--India program sponsored by the Indian
Space Research Organisation (ISRO).

\end{acknowledgement}

\begin{small}
\bibliographystyle{rr-assp}   
%%\bibliography{lahkar.bib}  

\end{small}
\end{document}